\documentclass[pra,twocolumn,showpacs,amsmath,amssymb,floatfix]{revtex4}

\usepackage{graphicx}
\usepackage{dcolumn}
\usepackage{bm}

\begin{document}

\title{Inducing elliptically polarized high-order harmonics from aligned molecules with linearly polarized femtosecond pulses}

\author{Adam Etches}
\author{Christian Bruun Madsen}
\author{Lars Bojer Madsen}
\affiliation{Lundbeck Foundation Theoretical Center for Quantum System Research, Department of Physics and Astronomy, Aarhus University,
8000 Aarhus C, Denmark}

\date{\today}

\begin{abstract}
A recent paper reported elliptically polarized high-order harmonics from aligned N$_2$ using a linearly polarized driving field [X.~Zhou \emph{et al.}, Phys.~Rev.~Lett.~\textbf{102}, 073902~(2009)]. This observation cannot be explained in the standard treatment of the Lewenstein model and has been ascribed to many-electron effects or the influence of the Coulomb force on the continuum electron. We show that non-vanishing ellipticity naturally appears within the Lewenstein model when using a multi-center stationary phase method for treating the dynamics of the continuum electron. The reason for this is the appearance of additional contributions, that can be interpreted as quantum orbits in which the active electron is ionized at one atomic center within the molecule and recombines at another. The associated exchange harmonics are responsible for the non-vanishing ellipticity and result from a correlation between the ionization site and the recombination site in high-order harmonic generation. 
\end{abstract}

\pacs{42.65.Ky, 42.65.Re}

\maketitle

\section{Introduction}
\begin{figure}
 \begin{center}
   \includegraphics[width=\columnwidth]{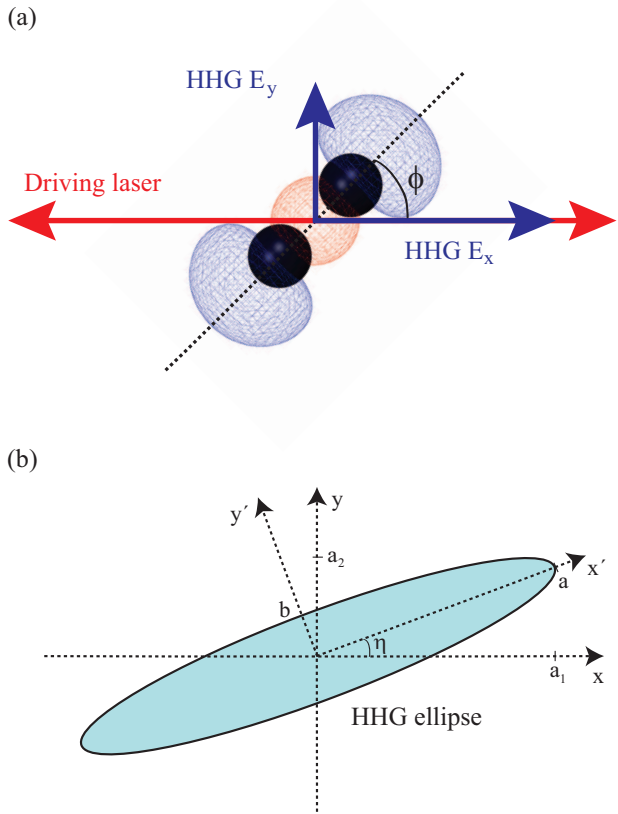}
 \end{center}
 \caption{(Color online) (a) The driving pulse propagates along the $z$-axis and is linearly polarized along the $x$-axis. The molecule is aligned at an angle of $\phi$ to the $x$-axis in the $xy$-plane. The alignment at a given angle with respect to the external field  results in harmonic emission with components $E_x$ and $E_y$. The hatched structure indicates the HOMO (cf.~Sec.~III). (b) The polarization ellipse for a single harmonic. The angle $\eta$ of the major axis and the length $b$ of the minor axis have been exaggerated for clarity.}
 \label{fig1}
\end{figure}

High-order harmonic generation (HHG) is a highly non-linear process in which a medium in an intense laser field emits coherent radiation at multiples of the driving frequency. The short time scale of the generation process makes HHG a promising source of coherent attosecond pulses in the XUV regime.
When the target is isotropic, such as a gas of atoms or unaligned molecules, it follows from symmetry arguments that the emitted harmonics have to be polarized parallel to the polarization axis of a linearly polarized driving laser. Breaking this isotropy by using a sample of aligned molecules for HHG allows for a non-vanishing perpendicular polarization component. This scenario is sketched in Fig.~\ref{fig1}. The perpendicular component is generally heavily suppressed compared to the parallel component. This has lead to a focus in the literature on the parallel component. Recent experiments on aligned N$_2$, O$_2$ and CO$_2$, however, have reported a non-vanishing perpendicular component~\cite{levesque:243001}. Moreover, elliptically polarized harmonics have been measured from aligned N$_2$ and CO$_2$~\cite{zhou:073902}. The presence of elliptically polarized harmonics opens up the possibility of generating elliptically polarized attosecond pulses in the XUV regime. 

From a theoretical point of view, the observation of elliptically polarized harmonics is very interesting because it serves as an important benchmark for different models. Several approaches are currently used to calculate the HHG response. Ideally, one should propagate the TDSE~\cite{gaarde:132001,lorin:025033}. This approach has been used for atoms and small molecules. An example is the prediction of comparable polarization components from aligned H$_2^+$ near minima in the spectrum~\cite{lein:183903}. For systems beyond H$_2^+$ and H$_2$ one commonly uses the semi-classical simple-man's model~\cite{corkum:1994} or the quantum mechanical Lewenstein model~\cite{lewenstein:2117} to describe HHG. Both models provide a three-step picture of the process in which a single active electron is brought into the continuum due to its interaction with the laser field, propagates in the field, and subsequently recombines and emits radiation. 

The calculation of HHG spectra using the Lewenstein model is generally combined with a stationary-phase method to reduce the computational cost. As discussed in Sec.~\ref{Standard stationary-phase method}, this procedure causes the resulting harmonics to be linearly polarized parallel to the polarization of the driving laser. This shortcoming has been attributed to the Lewenstein model itself~\cite{levesque:243001, zhou:073902}. However, the standard stationary-phase method is not in general valid for molecular targets as it fails to take into account the spatial extent of the molecule~\cite{chirila:023410}. Instead, an extended stationary-phase method can be applied to evaluate the Lewenstein model, which leads to so-called exchange harmonics~\cite{chirila:023410}. These are contributions to the total harmonic emission that are caused by quantum orbits describing an ionization event at one atomic center followed by recombination at another atomic center [see Fig.~\ref{fig2}]. 

We show that the Lewenstein model does lead to elliptically polarized high-order harmonics when exchange harmonics are included. Exchange harmonics rely on a correlation between the ionization and recombination events in HHG. This is different to the perpendicular harmonic component reported in~\cite{smirnova:063601}, which is obtained using scattering states for the continuum dynamics, thus causing a spread in momentum of the returning electron. The derivation in~\cite{smirnova:063601,ivanov:742} uses the standard stationary-phase method in order to separate the three steps of the HHG process. Consequently, any correlation, in the sense of exchange harmonics, between ionization and recombination sites is lost. Detailed comparison with experiments like that in~\cite{zhou:073902} will serve to determine the importance of this correlation.

The present paper is organized as follows. We briefly describe the Lewenstein model for aligned molecules in Sec.~\ref{The Lewenstein model}. The stationary-phase method is introduced in Sec.~\ref{Standard stationary-phase method}, and the resulting harmonics shown to be linearly polarized parallel to the linear polarization of the driving laser. In Sec.~\ref{Extended stationary-phase method} we present the extended stationary-phase method, and Sec.~\ref{Results and discussion} contains results demonstrating that such a  treatment of the Lewenstein model leads to elliptically polarized harmonics from aligned N$_2$. We summarize our findings in Sec.~\ref{Conclusion}. Atomic units $[\hbar = e = m_e = a_0 = 1]$ are used throughout unless stated otherwise.

\section{Theory}
\label{Theory}

We treat the response of a single molecule to the electric field $\mathbf F(t)$ of an intense femtosecond laser pulse. The laboratory frame is chosen such that the pulse propagates along the $z$-axis and is linearly polarized along the $x$-axis. We define the pulse by the vector potential
\begin{equation}
\label{A}
 \mathbf A(t) = \frac{F_0}{\omega_0} f(t) \cos(\omega_0 t) \mathbf e_x,
\end{equation}
where $F_0$ is the maximal field strength, $\omega_0$ the angular frequency, $f(t)$ the pulse envelope starting at $t=0$ and ending at $t=T$ (see Sec.~\ref{Results and discussion}), and $\mathbf e_x$ the polarization direction. The electric field is obtained from Eq.~\eqref{A} as ${\mathbf F} (t) = -\partial_t {\mathbf A} (t)$.

\subsection{The Lewenstein model for aligned molecules}
\label{The Lewenstein model}
The calculated spectrum $S_{\mathbf{n}} (\omega)$ of the harmonic component along $\mathbf n$ depends on the choice of gauge and form. Following~\cite{chirila:1039}, we use the length gauge for the interaction Hamiltonian, and the velocity form to determine the dipole acceleration. Thus, given the dipole velocity $\mathbf v_{\mathrm{dip}}(t)$, the signal is
\begin{equation}
 S_{\mathbf{n}} (\omega) = \left| \mathbf n \cdot \int_0^T dt \, e^{i\omega t} \frac{d}{dt} \left< \mathbf v_{\mathrm{dip}}(t) \right> \right|^2 \label{signal}.
\end{equation}
When evaluating Eq.~(\ref{signal}) it is necessary to include contributions from all occupied molecular orbitals $\lambda$~\cite{madsen:043419}, so 
\begin{equation}
 \left< \mathbf v_{\mathrm{dip}}(t) \right> = \sum_{\lambda} P_{\lambda} \left< \mathbf v_{\lambda}(t) \right>,
\end{equation}
with weights $P_{\lambda}$ fulfilling the normalization condition $\sum_{\lambda} P_{\lambda} = 1$. Introducing Euler angles $\mathcal R = (\phi, \theta, \chi)$ for the molecular orientation in the laboratory frame, the alignment distribution $G(\mathcal R)$ for a given molecule and alignment pulse can be calculated as in~\cite{ortigoso:3870}. Following~\cite{madsen:035401,madsen:043419}, this results in a coherent average over fixed alignment angles
\begin{equation}
 \left< \mathbf v_{\lambda}(t) \right> = \int d\mathcal R \, G(\mathcal R) \left< \mathbf v_{\lambda}(\mathcal R, t) \right>.
\end{equation}

The dipole velocity of a single perfectly aligned molecular orbital is calculated using the Lewenstein formula
\begin{eqnarray}
 \left< \mathbf v_{\lambda}(\mathcal R, t) \right> & = & i \int_0^t d\tau \, \int d\mathbf k \, {\mathbf v}^\star_{\mathrm{rec},\lambda}(\mathcal R, \mathbf k, t) \nonumber \\
 & & \times e^{-i S_{\lambda}(\mathbf k, t, t-\tau)} \mathbf F(t-\tau) \cdot \mathbf d_{\mathrm{ion},\lambda}(\mathcal R, \mathbf k, t-\tau) \nonumber \\
 & & + \text{complex conjugate} \label{Lewenstein}, 
\end{eqnarray}
where the nuclei are taken to be fixed during the driving pulse. This latter approximation is valid for heavy atoms such as nitrogen~\cite{madsen:023403,patchkovskii:253602}.

Equation (\ref{Lewenstein}) depends on the Fourier transform of the electronic wavefunction $\widetilde \psi_{\lambda}$ in both the ionization term
\begin{equation}
 \mathbf d_{\mathrm{ion},\lambda}(\mathbf k, t-\tau) = i \nabla_{\mathbf{k} + \mathbf{A}(t-\tau)} \widetilde \psi_{\lambda} [\mathcal R, \mathbf{k} + \mathbf A(t-\tau)] \label{d},
\end{equation}
and the recombination term
\begin{equation}
 \mathbf v_{\mathrm{rec},\lambda}(\mathbf k, t) = [\mathbf k + \mathbf A(t)] \widetilde \psi_{\lambda}[\mathcal R, \mathbf k + \mathbf A(t)] \label{v}.
\end{equation}
The vertical ionization potential $|E_{\lambda}|$ enters into the phase factor $S_{\lambda}(\mathbf k, t, t-\tau)$ as
\begin{equation}
 S_{\lambda}(\mathbf k, t, t-\tau) = \int_{t-\tau}^t \frac{1}{2} \left[ \mathbf k + \mathbf A(t'') \right]^2 dt'' - E_{\lambda} \tau.
\end{equation}

\subsection{Standard stationary-phase method}
\label{Standard stationary-phase method}

The momentum integral in Eq.~(\ref{Lewenstein}) describes the propagation of the active electron in the continuum. This integral is usually calculated using the stationary phase method, approximating it by a normalization factor times the value of the integrand at the stationary point $\mathbf k_{\mathrm s}$ of the rapidly varying phase.

Identifying the total phase with $S_{\lambda}(\mathbf k, t, t-\tau)$ gives the stationary-phase condition
\begin{equation}
 \nabla_{\mathbf k} S_{\lambda}(\mathbf k, t, t-\tau) \big|_{\mathbf{k}=\mathbf k_{\mathrm s}} = 0 \, , 
\end{equation}
which leads directly to the stationary momentum
\begin{equation}
 \mathbf k_s = -\frac{1}{\tau} \int_{t-\tau}^t \mathbf A(t'') \mathrm dt'' \label{k}.
\end{equation}
The standard stationary-phase method thus yields
\begin{eqnarray}
 \left< \mathbf v_{\lambda}(\mathcal R, t) \right> & \approx &  i \int_0^t d\tau \, \left( \frac{2\pi}{\epsilon + i \tau} \right)^{3/2}  {\mathbf v}^\star_{\mathrm{rec},\lambda}(\mathcal R, \mathbf k_{\mathrm s}, t) \nonumber \\
 & & \times e^{-i S_{\lambda}(\mathbf k_{\mathrm s}, t, t-\tau)} \mathbf F(t-\tau) \cdot \mathbf d_{\mathrm{ion},\lambda}(\mathcal R, \mathbf k_{\mathrm s}, t-\tau) \nonumber \\
 & & + \text{complex conjugate} \label{wrong approximate Lewenstein}.
\end{eqnarray}
Calculated spectra depend only weakly on the chosen value of $\epsilon$, which is set to $1$ in the following. 

Equations (\ref{k})--(\ref{wrong approximate Lewenstein}) select the dominant contribution to the full momentum integral in Eq.~(\ref{Lewenstein}). It is conventional to interpret the result of such a stationary phase analysis in terms of quantum orbits~\cite{salieres:05042001,faria:043407}. In this language, the active electron tunnels out, propagates parallel to the polarization of the laser, and recombines.

The vector character of (\ref{wrong approximate Lewenstein}) is determined by $\mathbf v_{\mathrm{rec},\lambda}(\mathcal R, \mathbf k_{\mathrm s}, t)$. Referring to Eqs.~(\ref{v}) and (\ref{k}), it is clear that the emitted harmonics are polarized parallel to the linear polarization of the driving laser. The Lewenstein model with the standard stationary phase method cannot give rise to ellipticities different from zero.

\subsection{Extended stationary-phase method}
\label{Extended stationary-phase method}
The treatment in Sec.~\ref{Standard stationary-phase method} is oversimplified in the molecular case~\cite{chirila:023410}. The reason for this is that $\mathbf d_{\mathrm{ion},\lambda}$ and $\mathbf v_{\mathrm{rec},\lambda}$ pick up phase factors related to the internuclear separation. This is readily seen when the electronic wavefunction is expanded around the atomic centers as
\begin{equation}
 \psi_{\lambda}(\mathbf r) = \sum_n \phi_n({\mathbf r} - {\mathbf R}_n) .
\end{equation}
The Fourier transforms in Eqs.~(\ref{d})--(\ref{v}) then separate into terms each relating to just one atomic center. This implies that
\begin{equation}
 \label{split d}
 \mathbf d_{\mathrm{ion},\lambda}(\mathbf k, t-\tau) = \sum_{n_i} \mathbf d_{\mathrm{ion},\lambda}^{n_i}(\mathbf k, t-\tau) ,
\end{equation}
and
\begin{equation}
 \label{split v}
 \mathbf v_{\mathrm{rec},\lambda}(\mathbf k, t) = \sum_{n_f} \mathbf v_{\mathrm{rec},\lambda}^{n_f}(\mathbf k, t) .
\end{equation}
%
%
Each term in Eq.~\eqref{split d}--\eqref{split v} acquires an associated phase factor $e^{i [\mathbf k + \mathbf A(t-\tau)] \cdot \mathbf R_{n_i}}$ or $e^{i [\mathbf k + \mathbf A(t)] \cdot \mathbf R_{n_f}}$, which has to be taken into account when performing the stationary-phase analysis. Here 
$n_i$ refers to the atomic center involved in evaluating $\mathbf d_{\mathrm{ion},\lambda}^{n_i}$, and 
$n_f$ refers to $\mathbf v_{\mathrm{rec},\lambda}^{n_f}$.

The above discussion means that Eq.~(\ref{Lewenstein}) can be split into 
\begin{equation}
 \left< \mathbf v_{\lambda}(\mathcal R, t) \right> = \sum_{n_i,n_f} \left< \mathbf v_{\lambda}^{n_i n_f}(\mathcal R, t) \right> \label{split Lewenstein}.
\end{equation}
The stationary-phase condition for $\left< \mathbf v_{\lambda}^{n_i n_f}(\mathcal R, t) \right>$ is
\begin{equation}
 \int_{t-\tau}^t \left( \mathbf k + \mathbf A(t'') \right) dt'' - (\mathbf R_{n_f} - \mathbf R_{n_i}) \Big|_{\mathbf{k}=\mathbf k_{\mathrm s}} = 0 ,
\end{equation}
with associated stationary momenta
\begin{eqnarray}
\label{ks_extend}
 \mathbf k_{\mathrm{s}}^{n_i n_f} = -\frac{1}{\tau} \int_{t-\tau}^t \mathbf A(t'') \mathrm dt'' + \frac{1}{\tau} \left( \mathbf R_{n_f} - \mathbf R_{n_i} \right) \label{k_st}.
\end{eqnarray}
Applying this multi-center stationary phase method to the integral in Eq.~(\ref{Lewenstein}) finally yields
\begin{eqnarray}
 \left< \mathbf v_{\lambda}(\mathcal R, t) \right> & \approx &  i \sum_{n_i n_f} \int_0^t d\tau \, \left( \frac{2\pi}{\epsilon + i \tau} \right)^{3/2} {\mathbf v}_{\mathrm{rec},\lambda}^{{n_f}^\star}(\mathcal R, \mathbf k_{\mathrm s}^{n_i n_f}, t) \nonumber \\
 & & \times e^{-i S_{\lambda}(\mathbf k_{\mathrm s}^{n_i n_f}, t, t-\tau)} \nonumber \\
 & & \times \mathbf F(t-\tau) \cdot \mathbf d_{\mathrm{ion},\lambda}^{n_i}(\mathcal R, \mathbf k_{\mathrm s}^{n_i n_f}, t-\tau) \nonumber \\
 & & + \text{complex conjugate} \label{approximate Lewenstein}.
\end{eqnarray}

The terms in Eq.~(\ref{approximate Lewenstein}) can be interpreted as semi-classical orbits of electrons that are ionized at $\mathbf R_{n_i}$ and recombine at $\mathbf R_{n_f}$. The effect of using the multi-center stationary phase method is to properly include orbits from one atom to another. The resulting harmonics are referred to as exchange harmonics~\cite{chirila:023410}. A similar idea was explored in an earlier paper~\cite{kopold:4022}, and its above threshold ionization analogue is well-known in the literature~\mbox{\cite{okunishi:043001,busuladzic:033412,busuladzic:203003}}.

The vector character of the emitted harmonics is determined by $\mathbf v_{\mathrm{rec},\lambda}^{n_i n_f}$ in Eq.~(\ref{approximate Lewenstein}). In analogy to the standard stationary-phase method, $\mathbf v_{\mathrm{rec},\lambda}^{n_i n_f}$ is parallel to $\mathbf A(t) + \mathbf k_{\mathrm s}^{n_i n_f}$. As seen from Eq.~\eqref{ks_extend}, this means that exchange harmonics will contribute with a non-vanishing perpendicular component if the molecular axis has a non-vanishing projection orthogonal to the driving laser. The conventional stationary phase method only includes the direct harmonics, which is why the standard treatment of the Lewenstein model fails to reproduce, even qualitatively, the findings of~\mbox{\cite{levesque:243001, zhou:073902}}. 

\begin{figure}
 \begin{center}
   \includegraphics[width=\columnwidth]{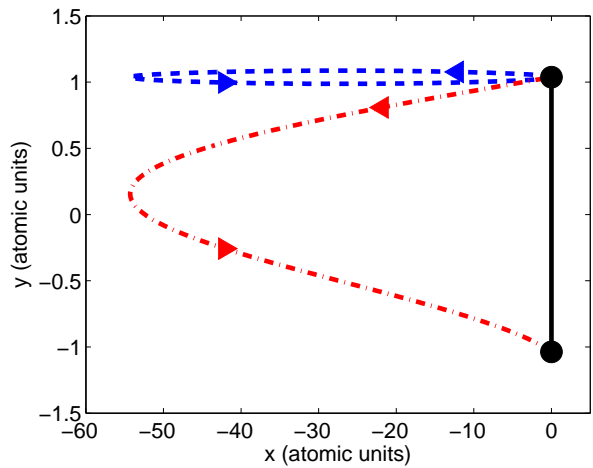}
 \end{center}
 \caption{(Color online) Two orbits with ionization at the peak of the field and recombination two thirds of an optical cycle later. The atomic centers of the molecule are sketched as black dots to the right. The laser polarization is taken to be in the $x$-direction. The dashed (blue) line represents an orbit that gives rise to direct harmonics, i.e., ionization and recombination at the same center. A transverse component has been added manually to distinguish the return path. The dash-dotted (red) line is an orbit that gives rise to exchange harmonics, i.e., ionization at one center and recombination at another center. Note that the units on the two axis differ.}
 \label{fig2}
\end{figure}

An example of the two types of orbits is presented in Fig.~\ref{fig2}. The electron is ionized at one atomic center at the peak of the field and recombines two thirds of an optical cycle later. Laser parameters are the same as those used in Sec.~\ref{Results and discussion} below. The orbit that gives rise to direct harmonics has been plotted with a small artificial transverse displacement in order to separate the two parts of the electron motion. Similar orbits exist which describe ionization at the other atomic center. 

We note in passing that the full Lewenstein expression in Eq.~\eqref{Lewenstein} is independent of whether a single-center or a multi-center expansion is adopted. The distinction is purely an artifact of using a stationary-phase method to approximate the momentum integral. In this setting, a multi-center expansion is preferable as it easily allows the geometric phase factors to be extracted.

\section{Results and discussion}
\label{Results and discussion}
\begin{figure}
 \begin{center}
   \includegraphics[width=\columnwidth]{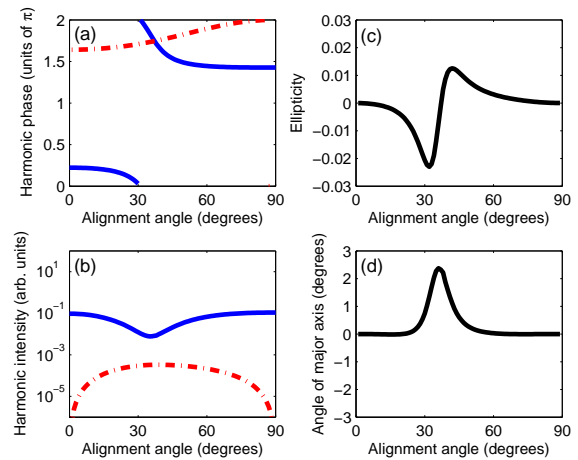}
 \end{center}
 \caption{(Color online) Orientation dependence of harmonic 17 from the 3$\sigma_g$ HOMO of N$_2$ with equilibrium internuclear distance $R_0=1.0977$~\AA. We use an $800$~nm, $6\times10^{14}$~W/cm$^2$ driving field. The envelope (for the vector potential) is trapezoidal with three optical cycles turn-on and turn-off and five cycles constant amplitude. (a)~Harmonic phase (b)~Harmonic intensity (c)~Ellipticity (d)~Angle of polarization ellipse major axis with respect to molecular axis. In (a) and (b) the solid (blue) curves refer to the parallel polarization component and the dash-dotted (red) curves to the perpendicular component. }
 \label{fig3}
\end{figure}

\begin{figure}
 \begin{center}
   \includegraphics[width=\columnwidth]{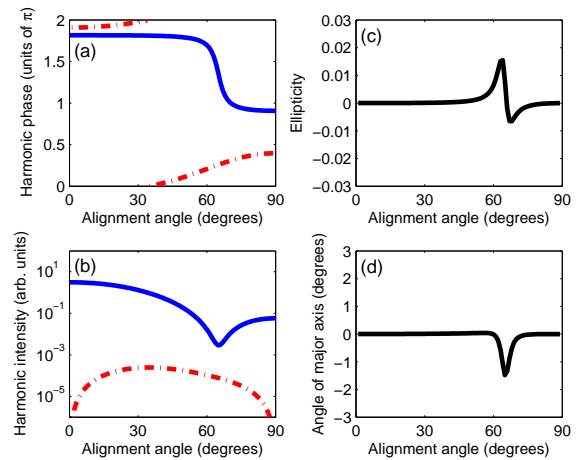}
 \end{center}
   \caption{(Color online) Same as Fig.~\ref{fig3}, but for harmonic 27.}
 \label{fig4}
\end{figure}

We present here a proof-of-principle calculation on N$_2$ in order to illustrate that Eq.~(\ref{approximate Lewenstein}) indeed leads to elliptically polarized harmonics when using a linearly polarized driving pulse. We assume only contributions from the $3 \sigma_g$ orbital, which is the highest occupied molecular orbital (HOMO). The HOMO is obtained using the standard quantum chemistry software package GAMESS--US~\cite{schmidt:1347} with a triple zeta valence basis set and diffuse $L$-shells. We assume perfect alignment, and perform no focal volume averaging. The alignment geometry and the polarization ellipse are shown in Fig.~\ref{fig1}.  Based on a wavelet analysis~\cite{antoine:r1750}, the $\tau$-integral in Eq.~(\ref{approximate Lewenstein}) is restricted to allow only short trajectories. This means that the upper bound is set to $0.67$ times an optical cycle. The driving field is taken to be an 800 nm pulse with 11 cycles in a trapezoidal envelope with 3 optical cycles for the linear ramp-up and ramp-down. The peak intensity is $6 \times 10^{14}$~W/cm$^2$.

We follow the polarization conventions in~\cite{born&wolf}. This means that the harmonic of angular frequency $\omega$ is assumed to be a perfect plane wave
\begin{eqnarray}
 \mathbf F^H(t) & = & a_1 \cos(\omega t + \delta_1) \mathbf e_x + a_2 \cos(\omega t + \delta_2) \mathbf e_y \\
 & = & a \cos(\omega t + \delta_0) \mathbf e_{x'} \pm b \sin(\omega t + \delta_0) \mathbf e_{y'},
\end{eqnarray}
where $\mathbf e_x, \mathbf e_y, a_1, a_2, \mathbf e_{x'}, \mathbf e_{y'}, a, b$ are defined in Fig.~\ref{fig1} (b). Phases $\delta_i$ ($i=0,1,2$) are found by assuming that the harmonic phase is equal to that of the dipole velocity~\mbox{\cite{lein:183903,chirila:013405}}.  
\begin{figure}
 \begin{center}
   \includegraphics[width=\columnwidth]{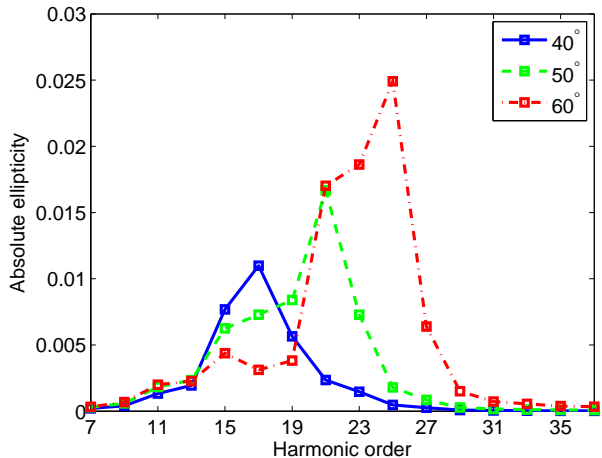}
 \end{center}
 \caption{(Color online) Absolute value of the harmonic ellipticity $\epsilon$ as a function of harmonic order for selected alignment angles $\phi$ of N$_2$. The nuclei are fixed at their equilibrium position. The solid (blue) curve refers to $\phi = 40^{\circ}$, the dashed (green) curve to $\phi = 50^{\circ}$, and the dash-dotted (red) curve to $\phi = 60^{\circ}$. See the caption of Fig.~\ref{fig3} for laser parameters.}
 \label{fig5}
\end{figure}\begin{figure}
 \begin{center}
   \includegraphics[width=\columnwidth]{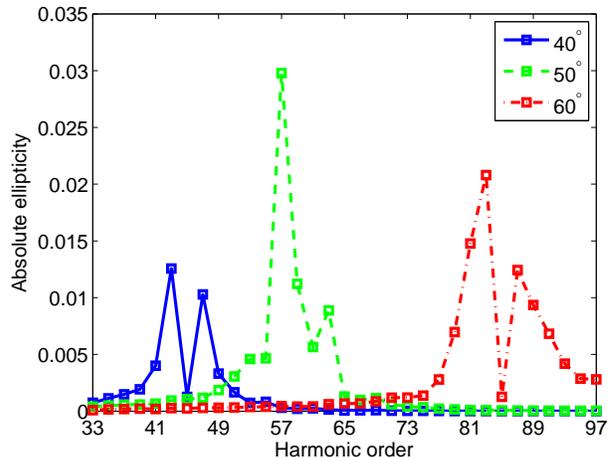}
 \end{center}
 \caption{(Color online) Same as Fig.~\ref{fig5}, but with the nuclei fixed at $R = 2R_0$.}
 \label{fig6}
\end{figure}

The two physically relevant quantities are the ellipticity $\epsilon$, and the angle $\eta$ that the polarization ellipse major axis makes to the polarization axis of the driving pulse. These are determined by
\begin{equation}
 \epsilon = \pm \frac{b}{a} \quad , \quad -\frac{\pi}{4} \leq \tan^{-1}(\epsilon) \leq \frac{\pi}{4} \label{epsilon}
\end{equation}
and
\begin{equation}
 \tan(2 \eta) = \frac{2 a_1/a_2}{1 - \left( a_1/a_2 \right)^2 } \cos(\delta_2 - \delta_1) \label{eta}.
\end{equation}

\subsection{Equilibrium nuclear spacing}
\label{Equilibrium nuclear spacing}

\begin{figure}
 \begin{center}
   \includegraphics[width=\columnwidth]{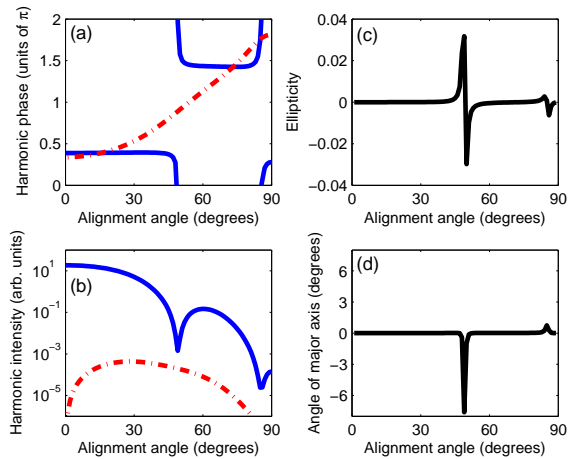}
 \end{center}
 \caption{(Color online) Same as Fig.~\ref{fig3}, but for harmonic 57 and with $R = 2R_0$.}
 \label{fig7}
\end{figure}

In this subsection we will restrict ourselves to the equilibrium nuclear spacing of $R_0 = 1.0977$ \AA. Figures~\ref{fig3}(a)--(b) and~\ref{fig4}(a)--(b) show data for harmonics 17 and 27, with solid (blue) curves referring to the component parallel with the linear polarization of the driving field and dashed (red) curves to the perpendicular component. Panels~(a) illustrate how the parallel polarization component of the harmonic emission changes its phase much faster than the perpendicular component. Comparison with panels~(b) shows that the rapid change in phase is accompanied by a drop in harmonic intensity, causing an increase in the ratio of perpendicular to parallel harmonic intensity.

We determine the ellipticity $\epsilon$ and the angle $\eta$ of the major axis using Eqs.~(\ref{epsilon})--(\ref{eta}). Results are shown in panels~(c) and (d). Both are seen to be very angle-dependent, taking non-vanishing values where the parallel component has a rapid change in phase. This variation in phase changes the helicity of the harmonics over a very short angle interval, while the angle of the major axis has a definite sign for a given harmonic.

Figure~\ref{fig5} shows the absolute value of the ellipticity as a function of harmonic order for selected alignment angles. The angular dependence can largely be understood by comparing panel~(b) in Fig.~\ref{fig3} with that in Fig.~\ref{fig4}. The dip in the intensity of the parallel component is fairly narrow and moves to larger angles for increasing harmonic order in agreement with the two-slit interference formula~\cite{lein:023805}. The resulting ellipticity does not compare well with~\cite{zhou:073902}. The experimental ellipticities are an order of magnitude higher and hardly have any angle-dependence.

The presented calculation shows that elliptically polarized high-order harmonics are predicted within the Lewenstein model. Improvements have to be made in order to compare the model directly with experimental data. One of these is to include the lower-lying molecular orbitals (HOMO-1, HOMO-2, etc.). The inclusion of these is expected to cause a dynamical minimum in the parallel component~\cite{smirnova:063601}, which would change the detailed structure. Taking into account the  distribution over alignment angles $G(\mathcal R)$ might also prove to be important.

\subsection{Extended nuclear spacing}
\label{Extended nuclear spacing}
As discussed above, exchange orbits give rise to elliptically polarized harmonics when the parallel polarization component is suppressed by the two-slit interference condition. Changing the internuclear separation will have an effect on this interference. Figure~\ref{fig6} shows results for N$_2$ with $R = 2R_0$. The plotted ellipticity is seen to peak at higher harmonic orders than was the case at $R = R_0$. There is no significant change in the magnitude of the obtained ellipticities. Figure~\ref{fig7} shows the detailed behaviour of the 55th harmonic. A second interference minimum appears at large alignment angles, which is consistent with the second order minimum in the two-slit model. 

\section{Conclusion}
\label{Conclusion}
Motivated by recent experiments~\cite{levesque:243001, zhou:073902}, we have addressed the issue of understanding how elliptically polarized harmonics arise in HHG when using a linearly polarized driving laser. It was shown that the Lewenstein model does not allow the emission of harmonics with a non-vanishing perpendicular component if the standard stationary-phase method is used to evaluate the continuum dynamics. 
A multi-center stationary phase method was adopted in order to take into account the spatial extent of the molecule. This leads to additional terms in the Lewenstein model compared to the standard stationary-phase method. These were interpreted as quantum orbits describing an ionization event at one atomic center followed by propagation in the continuum and recombination at a different atomic center~\cite{chirila:023410}. We illustrated that the total harmonic emission from N$_2$ calculated using this method is elliptically polarized near minima in the spectrum. 

The electron continuum was described using only Volkov waves, as opposed to the use of scattering states~\cite{smirnova:063601}, indicating that the appearance of elliptically polarized harmonics is in part the result of having an extended target. Further work is required to ascertain the importance of the Coulomb interaction between the active electron and the molecular ion.

\section*{Acknowledgements}
We thank Brett D.~Esry for fruitful discussions. This work was supported by the Danish Research Agency (Grant No.~2117-05-0081).

\end{document}